\documentclass[epj]{svjour}
\usepackage{url}
\usepackage{graphicx}
\usepackage{amssymb}
\title{Boom and bust in continuous time evolving economic model}
\author{L. Mitchell\thanks{E-mail: \email{lawrence.mitchell@ed.ac.uk}}
 \and G. J. Ackland\thanks{E-mail: \email{g.j.ackland@ed.ac.uk}}}
\institute{SUPA, School of Physics, The University of Edinburgh, Mayfield
  Road, Edinburgh EH9 3JZ, United Kingdom}

\abstract{We show that a simple model of a spatially resolved
  evolving economic system, which has a steady state under
  simultaneous updating, shows stable oscillations in price when
  updated asynchronously.  The oscillations arise from a gradual
  decline of the mean price due to competition among sellers competing
  for the same resource.  This lowers profitability and hence
  population but is followed by a sharp rise as speculative sellers
  invade the large un-inhabited areas. This cycle then begins again.
}
\begin{document}

\maketitle{}
\section{Introduction}
Cycles in natural and economic systems are widely observed and are
often described by analogy to the differential equations describing
physical oscillating systems. However, the observed oscillations are
frequently irregular.  In some cases, such as the Milankovitch cycles
which drive ice ages, this irregularity may be an interplay between
different deterministic modes (precession, obliquity and
eccentricity), however the possibility of stochastic effects producing
low frequency oscillations has been widely overlooked.

In economics, Schumpeter's equilibrium theories related business
cycles to oscillations about equilibrium in a dynamical system. Other
authors relate cycles to time lags in the system: the cobweb model
\cite{Kaldor:1934} in which price is determined by supply, which is in
turn determined by previous prices; or Kitchin's inventory model which
has prices determined by stored goods \cite{Kitchin:1923}.  See also
the macroeconomic models of Lucas \cite{Lucas:1975} and Blinder
\cite{Blinder:1981}.  Further authors have proposed bubbles of
overconfidence (e.g.~Elliot \cite{Frost:1995}, Juglar) or external
controlling influences (as, for example, governmental policy changes)
as the driving force behind business cycles. Finally, Kondratieff's
long wave cycle \cite{Kondratieff:1935} has cycles driven by exogenous
innovation shocks.  We might imagine that these should be irregularly
spaced and stochastic in nature, they are not however treated as such.
This is in part because the measured quantities (various economic
indicators) are seen as separate from the underlying driving forcer.

In ecological systems, the model of Lotka-Volterra gives regular
oscillations, while more recent food-web and evolutionary models
\cite{Bak:1993,Laird:2007} tend to have stable epochs punctuated by
rapid change in the spirit of the punctuated equilibrium of Eldredge
and Gould \cite{Eldredge:1972,Gould:1977}.  Regular oscillations
driven by lags in reacting to a change in environment are also
observed \cite{Bascompte:1997} (in a similar manner to the
above-mentioned economic models).

\section{A simple marketplace model}
In a previous paper, we introduced a simplified model of a market
subject to none of these drivers, rather relying on supply-led
competition and evolution through bankruptcy and startups
\cite{Mitchell:2007}.  Competition occurs through price -- low prices
increase the chance of sales, but reduce the profit margin.  A dual
ecological model considers competition for resources -- effective
foraging requires a high metabolic rate and a need to eat more
frequently.  In this paper we will use the language of the economic
model.

Here we consider a continuous time version of the model, and show that
it generates spontaneous oscillations of irregular period.

\begin{figure}[htbp]
  \includegraphics[width=8.8cm]{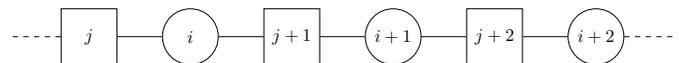}
  \caption{Diagram showing lattice of buyer-seller interactions.
    Sellers are marked as squares, buyers as circles.}
  \label{fig:1}
\end{figure}

The details of the model are as in previous work \cite{Mitchell:2007}
except for the choice of update scheme.  We consider a ring of $2N$
alternating sellers and buyers ($N$ of each), the lattice layout
showing connections is shown in Fig.~\ref{fig:1}.
  Each seller
has capital $C_i$ and fixed price $P_i$ ($i=1,\dots,N$).  Initial
prices are drawn from $P_i \in [0, P_{\textrm{max}})$, and initial
capital is $C_i = 0 \; \forall i$.  A timestep consists of N
iterations of the following:
\begin{enumerate}
\item One seller ($j$) is chosen uniformly at random and pays overhead
  $d=2$, decreasing the capital $C_j$ by 2.
\item One buyer ($i$) is chosen uniformly at random and the lower
  priced of the two adjacent seller sites ($k = j\textrm{ or }j+1$)
  increases its capital $C_k$ by $P_k$.
\end{enumerate}
Hence in each timestep, on average, each buyer makes one purchase and
each seller pays one overhead.

Sellers which have not paid an overhead still have the ability to
produce goods for a buyer: we treat the overhead as a fixed cost,
rather than a payment for the acquisition of goods to sell. At the end
of the timestep, we now remove bankrupt sellers in the following
manner:
\begin{enumerate}
\item All sellers with $C_i < 0$ are bankrupt: site $i$ becomes vacant.
\item Vacant sites are repopulated with probability $\gamma$.
\item New sellers at site $i$ have $C_i = 0$ (writing off any debt the
  previous seller may have had).
\item New sellers at site $i$ take the price of a randomly chosen
  existing site $j$, $P_i = P_j + dp$, ($dp
  \in[-\min(\Delta, P_j), \Delta]$).
\end{enumerate}
This completes one complete timestep for the model.  Note that the
buyer sites are always occupied but seller sites may be vacant: in
some rounds, some buyers have no choice of supplier.  The only two
free parameters in the model are $\gamma$ and $\Delta$.

It is worth mentioning that the seller dynamics mean that sellers have
neither rational expectations of the market, nor any explicit memory
of past transactions: implicitly the market behaviour is encoded in
the time evolution of seller capital, this is however not consulted
when choosing a strategy except for the case of bankruptcy when
$C(t-1) > 0$ and $C(t) < 0$ when the strategy of the seller changes.
The hold-over of capital by sellers is similar to a model of Lucas
\cite{Lucas:1975} in which it is found that exogenous unanticipated
shocks to a collection of independent markets (Phelpsian Islands) can
have the effect (if there is memory causing information lags) of
producing cyclical output.  In our model there are no external shocks,
the variations in the market being driven solely by the internal
dynamics.

\begin{figure}[htbp]
  \includegraphics[width=8.8cm]{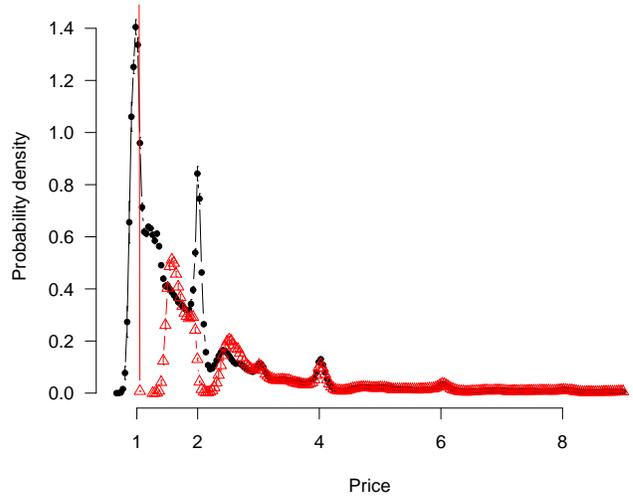}
  \caption{Comparison of steady state price distribution for
    continuous (black, circles) and discrete (red, triangles) time
    simulations with $N=10^6$, $\Delta = 0.04$, $\gamma = 0.5$.  Error
    bars show standard error in the mean for 20 time-averaged
    distributions sampled 1000 timesteps apart after steady state is
    reached.  Note that for the continuous time model, there is no
    fixed break even price, although on average $P=1$ is break even.
    High prices cut off at $P=9$ for clarity, along with the discrete
    time peak at $p=1$ which reaches 12.}
  \label{fig:2}
\end{figure}

For high values of the rebirth parameter $\gamma$, only sellers
charging close to their marginal cost can survive, however for $\gamma
\lessapprox 0.5$ the price distribution formed favoured peaks
\cite{Mitchell:2007} at higher prices.  We will refer to sellers with
$P_i\approx 1$, $\langle C_i \rangle \approx 0 $ as `cheap' and those
with sufficiently high price to accumulate capital $\langle C_i\rangle
> 1$ `expensive'.  We find that expensive sellers are still present in
the continuous time formulation as shown in Fig.~\ref{fig:2}.

Note that unlike many microscopic models
\cite{Shnerb:2000,Guan:2006,Nowak:1994} of evolutionary competition,
the favouring of certain strategies is not enhanced by niche
construction for similar organisms.  Rather, sellers are only able to
construct niches which enhance the selection of competitor strategies.
Expensive sites do not survive by conferring an advantage on similar
expensive sites (the cheap sellers have a much higher fitness when
invading): niches for expensive sellers exist in gaps between cheap
sites \cite{Mitchell:2007}.

\section{Oscillatory behaviour}
\label{sec:transient-dying-out}

\begin{figure}[htbp]
  \includegraphics[width=8.8cm]{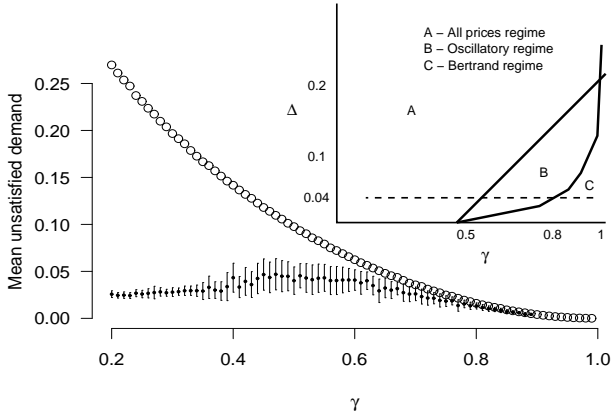}
  \caption{Ensemble-averaged unsatisfied demand (fraction of unsold to
    buyers) as a function of birth-rate, $\gamma$, in an enforced
    Bertrand steady state (open circles, error bars too small to show)
    and unrestricted steady state (solid circles), $N=10^4$, $\Delta =
    0.04$.  Error bars show standard error in the mean over thirty
    ensembles after steady state is reached.  In contrast to the
    discrete time version, the Bertrand solution diverges from
    unrestricted result at $\gamma \approx 0.8$
    (c.f.~\cite{Mitchell:2007} where the two are identical until
    $\gamma \approx 0.6$).  This intermediate regime ($0.5 \lessapprox
    \gamma \lessapprox 0.8$) is the region of oscillatory behaviour.
    Inset shows a schematic phase diagram with the three regions
    marked, dotted line shows position of in phase space of
    unsatisfied demand data.}
  \label{fig:3}
\end{figure}

As $\gamma$ is increased above $\approx 0.5$, the system enters an
oscillatory state after an early time transient decay (erasing the
initial conditions).  The phase diagram (Fig.~\ref{fig:3} insert)
shows schematically the region in parameter space which leads to such
a situation.  These oscillations are evident as a slow decrease
followed by a fast increase in the total number of sellers and the
mean price of the system.  This boom-bust cycle is stable across many
oscillatory periods and is occasionally interspersed with larger
excursions.

This regime is not seen in the discrete time model
\cite{Mitchell:2007} since the high-priced sellers are not able to
persist in unfavourable situations.  Stochastic dynamics can however
allow their persistence, since it is not a given that an expensive
seller will have to pay overhead during a round in which it is not
selling.  This is enough to persist long-lived expensive sellers,
which as we shall see drive the oscillations.

\subsection{Aperiodicity}
\label{sec:aperiodic-oscillations}

\begin{figure}[htbp]
  \includegraphics[width=8.8cm]{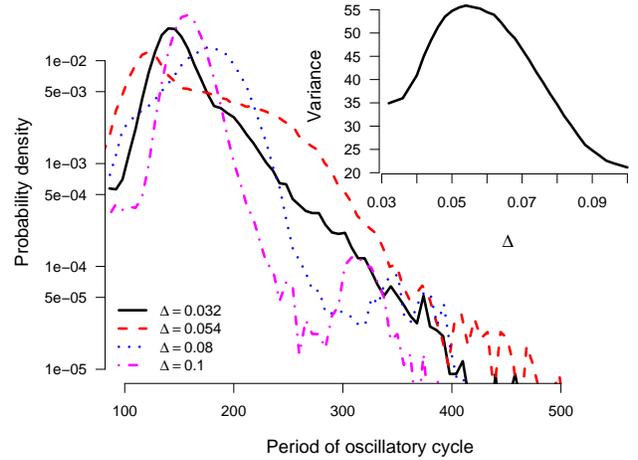}
  \caption{Distribution of cycle period lengths for different values
    of $\Delta$ as indicated in legend, $N=10^4$, $\gamma = 0.7$.
    Data prepared by smoothing time-series for the number of live
    sellers with a 40-point moving average to remove high-frequency
    components and then calculating the distance between zeroes of the
    derivative of the dataset.  The peak and left tail are well-fitted
    by Normal distributions while the right tail is heavier (with an
    approximate exponential decay).  Inset shows how the variance of
    the distribution varies with $\Delta$.}
  \label{fig:4}
\end{figure}

Unlike other models of boom-bust cycles (see for example
\cite{Mullineux:1993} for a review), we find that the period of the
oscillations is not a simple function of the model parameters,
instead, the period is stochastic in nature with a typical length
(Fig.~\ref{fig:4}), the origin of this stochasticity is explained in
greater detail below.  The shape of the distribution takes broadly two
forms, for `small' $\Delta$ the peak is Gaussian with a heavier,
exponential right tail of high prices.  For `large' $\Delta$ more of
the probability mass is in the peak with a less pronounced right tail.
These differences may be explained by considering the exact character
of the oscillations.  When $\Delta$ is small ($\Delta \lessapprox
0.05$) the range of prices that the oscillations cover is typically
also small $0.6 < \bar{P} < 1.3$ with intermittent larger excursions
to $\bar{P}\approx 2$.  For larger $\Delta \gtrapprox 0.07$, the
typical oscillation is over a wider range of price $0.08 < \bar{P} <
2$ with fewer large or small excursions.  The modal period of these
differing regimes is approximately the same as $\Delta$ also
influences the rate of change of $\bar{P}$. This compensates to some
extent for the larger change required in $\bar{P}$ to complete a
period.  The cross-over between the two regimes is gradual which may
be seen by considering the variance of the period distribution as a
function of $\Delta$ (Fig.~\ref{fig:4} inset).

\section{Dynamics of the oscillations}

\begin{figure}[htbp]
  \includegraphics[width=8.8cm]{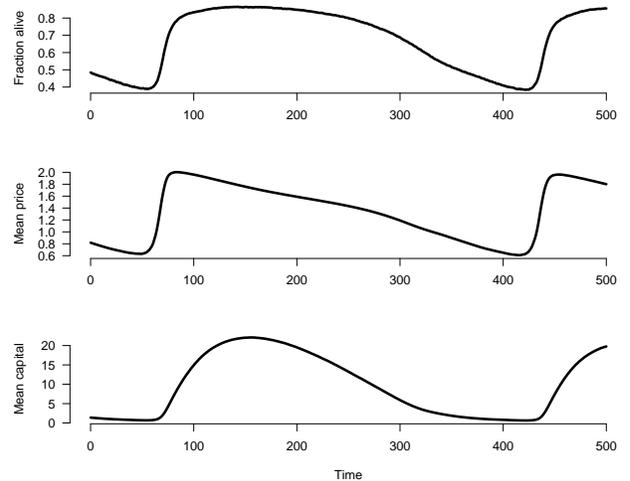}
  \caption{One period of oscillations for a system with $N=10^5$,
    $\gamma=0.75$ and $\Delta = 0.04$.  Shown from top to bottom are
    the fraction of live sellers, the mean price of those sellers and
    the mean capital.  Note how the mean capital peaks later than
    the price and grows more slowly}
  \label{fig:5}
\end{figure}

In the oscillatory regime, the mean price typically fluctuates between
$\bar{P} < 1$ and $\bar{P} \approx 2$ with a sharp upswing and slow
decay as shown in Fig.~\ref{fig:5}.  The asymmetry in the cycle may be
explained by considering the different mechanisms involved in the
upswing and downturn.  The driving force on the downward section of
the oscillation is competition between many like-priced sellers.  In
this case, a lower price is favourable.  The upswing, however, is
nucleated from a few existing high-priced sites: rather than evolution
from cheap sites, a peak suddenly appears in the price distribution
which then moves slowly downwards.  The plateau at the top of the
oscillation is due to a decreased rate of seller turnover: when all
competing sellers have a similar price, mean lifetime of sellers
increases with the price, hence, at the top of a cycle, the turnover
rate will be low causing a `slowing' of the dynamics.

This asymmetry in the oscillatory waveform leads to another
interesting result.  We find that as the amplitude and period of the
oscillation increases, the mean fraction of live sellers over a period
also increases, along with the total number of sales.  This is easily
understood given the previous argument, the system spends more time at
the top of the cycle (since seller turnover rate is slower) than at
the bottom.  We note that this behaviour is similar to that of maximum
shown in \cite{Wood:2006} for a spatial Daisyworld model. another type
of evolving stochastic cellular automaton.

We may explain the oscillatory cycle by considering the variation in
the fraction of vacant sites over a cycle.  As the mean price drops,
the demand in the system is able to support fewer and fewer sellers:
for a mean price $\bar{P}$, the demand in the system supports
approximately $\bar{P}/2$ sellers.  The increasing number of vacant
sites allows expensive sellers to thrive in locally favourable
environments: if a buyer is next to a vacant site, it has to visit the
seller on the other side of it (even if this seller is high-priced).
These expensive sellers survive to the end of the round and accumulate
sufficient capital to survive even if they fail to make a sale in the
following round.  They are thus candidates for replication into newly
vacant sites.  Since only live sellers are considered during the
replication cycle, the expensive sellers occupy the space -- they can
survive a timestep without a sale because one sale can cover multiple
overhead payments.  As the system fills up with high-priced sellers,
new sellers in vacant sites do better by undercutting existing ones,
rather than exploiting vacant sites of which there are now few.  The
price of the system is then driven downwards by Bertrand competition
\cite{Bertrand:1883}.

\section{Persisting sellers}
\label{sec:ancestry}

As already intimated, the upswing in the price oscillation has a
different origin to the downturn: Bertrand competition among sellers
favours $P=1$.  We may track this more closely by observing the
ancestry of sellers through time.  We label each seller uniquely at
the beginning of a simulation, every new seller adopts its label and
price from a parent seller.  This allows us to associate sellers with
a given `franchise'. We first note that in the oscillatory regime the
number of unique ancestors falls to $O(1)$ after a finite time
(exponential decay).  In contrast, in the stationary regime, the
number of unique ancestors decays with an approximate power law with
exponent close to unity. Thus, sellers do not collapse onto a single
common ancestor, but rather a number of different franchises persist.

\begin{figure}[htbp]
  \includegraphics[width=8.8cm]{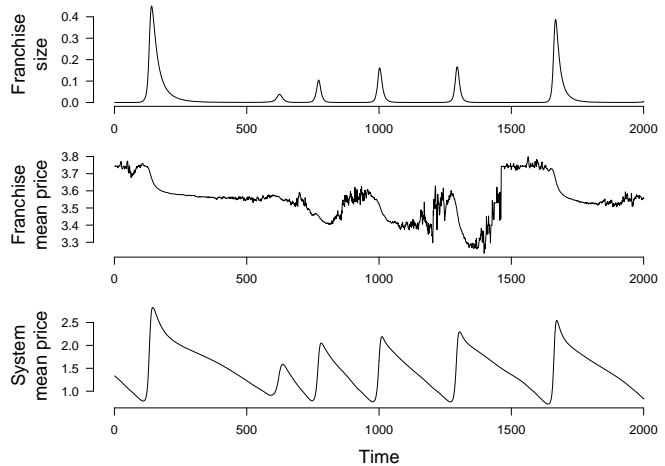}
  \caption{Change in fraction of sellers with a common ancestor (a
    franchise) through a number of oscillatory cycles. $N = 10^5$,
    $\gamma = 0.7$, $\Delta=0.08$.  Top panel shows the size of the
    franchise as a fraction of total sellers.  Middle panel shows the
    mean price of sellers in the franchise.  Bottom panel shows the
    mean price of the system.  The original parent persists through
    many cycles while the offspring are only short-lived, flourishing
    during upswings in the mean price.}
  \label{fig:6}
\end{figure}

We now look for franchises with a high mean price: expecting that
during the bottom of a cycle they will have only a few members; and
during an upswing their size should increase markedly.  As
Fig.~\ref{fig:6} shows, this is indeed the case: in the oscillatory
regime we find a small number of persisting high-priced franchises.
During lean periods, these franchises often only consist of a single
seller.  These sellers are the parents of a franchise.  They have been
able to build up a large capital which allows survival through bad
patches when they are unable to sell.  Offspring sellers, however,
have less accumulated capital and are thus less likely to survive
competition with cheaper sellers once Bertrand competition kicks in.
Hence, typically only the original parent (and perhaps a few
offspring) survive without selling until the bottom of the cycle is
reached.

The downward change in mean price arises from a contribution of
pre-existing franchises flourishing at different times and adaptation
within some franchises to the current fittest strategy.  Very
high-priced franchises ($P \gtrapprox 3$) do not seem to adapt much
and thus their contribution to the mean price is only due to varying
size.  Intermediate ($P \approx 2$) and cheap franchises both vary
their size and adapt to the current fittest strategy.  We can see this
occuring in Fig.~\ref{fig:6}.  During the third cycle the mean price
reaches $2.75$ and initially falls off quite sharply: sellers jumping
ship from one franchise to another.  As the mean price reaches $\sim
2$ the franchises start adapting and the rate of change decreases.

Concurrently persisting franchises have slightly different
prices. Thus the maximum mean price over a period is determined by how
successful each of the franchises are in invading the empty sites at
the bottom of a cycle.  Since both the rate of change of $\bar{P}$ and
its minimum are approximately constant for a given $\Delta$, the
observed distribution of periods is primarily due to the stochastic
manner in which franchises achieve success (invasion) at the beginning
of a cyle.

This oscillatory behaviour can be interpreted in terms of the fitness
of the sellers - defined as their number of offspring.  The fittest
price is environmentally determined, depending on the number of vacant
sites.  A new expensive mutant is unable to invade a system
nearly-full of cheap sellers unless it is located adjacent to a vacant
site (and therefore has a buyer).  If the number of vacant sites
increases, the mutant's fitness increases (depending on its position)
and it is able to survive and proliferate.  This newly established
mutant-rich system is now invadable by cheaper sellers.  Thus, the
system is found to be in an oscillating state alternately trying to
fix to a population of cheap sellers (which are fitter when the
predominant sellers are expensive) and a population of expensive
sellers (fitter when the predominant sellers are cheap).

\section{Suppressing oscillatory behaviour}
\label{sec:supr-oscill-behav}

In this section we study ways of avoiding the system-wide oscillatory
behaviour while allowing for the continued existence of highly
profitable sellers.  The first method we study divides the system into
a number of large, semi-autonomous, regions.  We have seen that the
oscillatory cycles the system goes through do not have a fixed period
and hence, if we divide the system into regions, they will be likely
to oscillate out of phase with one another.  The net result on the
global mean price will be a reduction in the amplitude of observed
oscillations.  Obviously, if the regions are completely independent,
we have not gained anything, just shown that out of phase oscillations
can partly cancel.  Consider, however, if we couple the regions
somehow.  If the coupling is weak, the regions may continue out of
phase.  Recall that the upswings are seeded by just a few high-priced
sellers.  Should all of these sellers die out, the system price can
never recover and consequently the mean price stays low.  If this
system were weakly coupled with another oscillating region, an
expensive price could be copied in from outside, reseeding the
expensive franchises and allowing for recovery to high prices.  With a
large number of regions, the likelihood that they all simultaneously
crash to the low-priced state is very low.  This should allow for
longer survival times of an oscillating state over similarly-sized
systems without separate regions.

\begin{figure}[htb]
  \includegraphics[width=8.8cm]{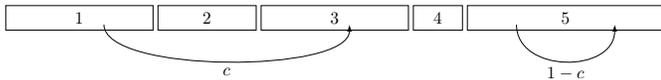}
  \caption[Island divisions in one dimension]{Diagram showing the
    division of one-dimensional system into islands.  This particular
    system has five islands (labelled).  Also shown are two copying
    events: an inter-island copy (occurring with probability $c$) and
    an intra-island copy (probability $1-c$)}
  \label{fig:7}
\end{figure}
\begin{figure}[htbp]
  \includegraphics[width=8.8cm]{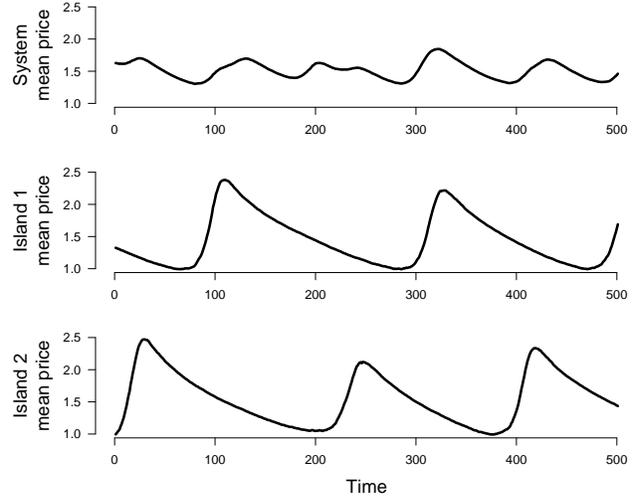}
  \caption{Mean price exhibited in an oscillating system divided into
    five equally sized islands.  Top figure shows the mean system
    price, bottom two figures show the mean price of two different
    islands.  Note how the amplitude of the oscillations is much
    reduced for the global system due to out of phase oscillations
    between islands.  $N=10^5$, $\Delta = 0.04$, $\gamma = 0.75$,
    $c=0.05$}
  \label{fig:8}
\end{figure}

The coupling we introduce is to divide our system into regions.
Buyers do not see the region boundaries, however, sellers do.  When a
new seller is introduced into the system they either take their price
from within their region, or from the whole system.  The former occurs
with some set probability $1-c$, the latter with probability $c$.
Taking $c=1$ corresponds to the previously studied case, $c=0$ leads
to completely uncoupled regions.  Figure \ref{fig:7} shows a
diagram of the two different copying steps.  We find that for small
values of the coupling constant ($c \lessapprox 0.05$), islands
oscillate out of phase with on another.  This leads to a stabilisation
in the mean price exhibited by the whole system
(Fig.~\ref{fig:8}), but without affecting the dynamics in
each individual island.  
With $c \gtrapprox 0.05$, the separate islands oscillate in phase and
the system then evolves as if the island separation did not exist
(i.e., the global oscillations are the same as the island ones).  The
low coupling regime allows for a recovery from crashes that the latter
does not.  Since the upswings are driven by a small number of
expensive sellers, it is possible that all these sellers become
bankrupt, the system cannot recover from this crash.  When islands
oscillate in phase, this crash is global.  With out of phase
oscillations, other islands can be in a high-priced phase of the
cycle, these sellers can then reseed the crashed part of the system
with new expensive sellers, leading to a better recovery.

The second method of suppressing oscillations we study is to introduce
a varying time-delay in the price copying stage.  We can think of this
as a random delay in the propagation of news events.  The driving
cause of the downward phase of the cycle is Bertrand competition
between sellers favouring cheaper prices.  The repeatability of the
cycles indicates that there is only one possible route to the
low-priced state the dynamics can take.  By destroying (to some
degree) the correlation between the copied price distribution and the
exhibited system price distribution, we are able to suppress this
pathway, removing the oscillatory cycle.

\begin{figure}[htbp]
  \includegraphics[width=8.8cm]{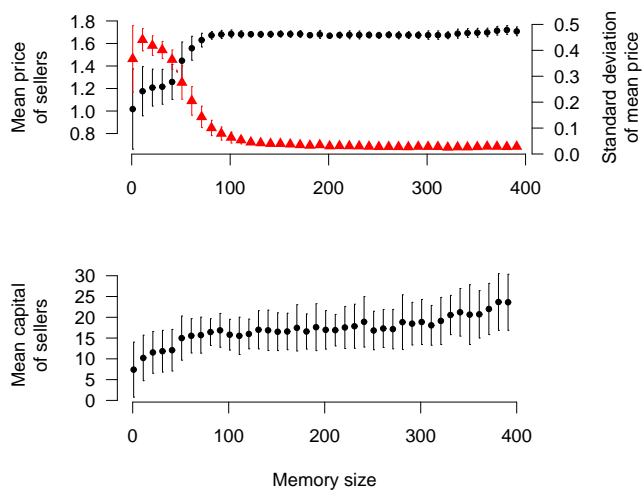}
  \caption{Effect of varying the memory size on system prices.  The
    top figure shows the mean price of sellers (left axis, black
    circles) over 25000 rounds after equilibriation, and the standard
    deviation around that value (right axis, red triangles).
    Errorbars show standard error in the mean over forty realisations.
    Increasing the memory size reduces fluctuations in the mean price
    and increases its equilibriated value.  Lower figure shows the
    mean capital of these sellers: increasing the memory size weakly
    increases the mean capital.  $N=10^5$, $\Delta=0.04$, $\gamma =
    0.75$.}
  \label{fig:9}
\end{figure}
To remove this correlation, we allow each seller to retain a memory of
its price over some fixed number of selling rounds.  When copying a
price, a seller to copy is picked and then a price is chosen uniformly
at random from the set of available historical prices.
Fig.~\ref{fig:9} shows the effect of varying the memory size on the
system mean price.  We find for short memories (up to around 30
rounds), the oscillatory behaviour is retained.  As the memory gets
longer, the oscillations are suppressed until eventually the system no
longer displays any oscillatory behaviour.  Interestingly, this change
to the dynamics recovers the structure (though not the exact form) of
the steady state obtained for low $\gamma$.  This change to the
dynamics increases both the mean price and the mean capital in the
system.  Additionally, the profitability of anomalous sellers (those
with huge capital buildups) is unaffected by the change.  These
sellers (which are the parents of expensive franchises in the
oscillatory phase) are still able to obtain a large capital.  If we
wish to optimise our system for overall prosperity, larger memories
are better.

\section{Conclusions}
\label{sec:conclusions}

We have seen that modification of the discrete time model presented in
\cite{Mitchell:2007} to a continuous time version can lead to the
appearance of self-sustaining boom-bust cycles.  The cycle arises from
continual pressure to reduce prices by Bertrand competition, until the
price becomes unsustainable and only sellers with large accumulated
capital survive.  These expensive sellers proliferate when there are
many vacant sites, until Bertrand competition between them drives the
modal price down once more.

There is no externally-imposed timescale for oscillation, the typical
period emerges from the internal dynamics with some variation from
cycle to cycle.  The oscillations observed show that in order to model
the `stylised fact' of asymmetric business oscillations
\cite{Artis:2004,Sichel:1993,Neftci:1984} it may be enough to require
stochasticity in the number of sales (and a separation of timescales
between selling and bankruptcy) in a spatially separated market.
Rather than require exogenous innovation shocks, our model requires
only a variability in the amount of competition expensive sellers
experience.  Assuming they can survive depression periods, these
sellers are able to exploit a favourable market when one such arrives.
We note, however, that our model produces cycles which consist of
rapid upturns followed by a slower decay: this is in contrast to the
asymmetries typically observed in real world business cycles for which
the decay is rapid and the upturn more gradual.  The behaviour seen is
perhaps more similar to that observed during introduction of new
technologies \cite{Solomon:2000}, especially when considering the mean
cpaital.  Initial high prices and high profitability followed by a
gradual decline of price as the market catches on.  Finally the new
product becomes old and is again superseded.

The length of each oscillation is primarily determined by the price of
the expensive franchise which first becomes established after the
bust phase.  The higher this is, the longer it takes for Bertrand
competition to take effect.  This is a purely stochastic effect,
qualitatively different from previous models of boom and bust.

We have studied two methods of stabilising the system against
oscillations.  Dividing the system into parts which oscillate
independently reduced the observed effect on the global behaviour and
allowed reseeding of expensive sites from crashed systems.  We found
that adding a random delay to the price copying stage completely
removed the oscillatory behaviour and the system returned to a steady
state with a price distribution similar to that found at low $\gamma$
values.

From an evolutionary point of view, we divide the sellers into species
by ancestry. We find that the change in mean price arises from a
combination of pre-existing species flourishing at different stages of
the cycle, along with inter- and intraspecies Bertrand competition.
The former is the primary driver during upturns, while downturns are
due to a combination of both drivers.  More concretely, very
high-priced franchises do not participate in Bertrand competition.
Mid- and low-priced franchises do, which causes the slow downswing.
Thus, in the ecological context, although we have single-parent
(haploid) organisms, the inherited ancestry label (genotype) tends to
correspond to a consistent price (phenotype) and so it makes some
sense to regard the model as producing distinct species with high and
low prices.

\begin{acknowledgement}
  This work was produced by the NANIA collaboration funded by EPSRC
  grant T11753.  We thank three anonymous referees for useful
  comments.
\end{acknowledgement}

\end{document}